\documentclass{webofc}

\usepackage[varg]{txfonts}   
\usepackage{hyperref}
\usepackage{url}
\usepackage{graphicx, subfigure}
\hypersetup{colorlinks=true,citecolor=blue,urlcolor=blue,linkcolor=blue}
\begin{document}
\title{Hydrodynamization and thermalization in heavy-ion collisions: a kinetic theory perspective}
%
%

\author{\firstname{Caio} \lastname{V.~P.~de Brito}\inst{1}\fnsep\thanks{\email{caio_brito@id.uff.br}} \and
        \firstname{Gabriel} \lastname{S.~Denicol}\inst{1}\fnsep\thanks{\email{gsdenicol@id.uff.br}} 
}

\institute{Instituto de F\'{\i}sica, Universidade Federal Fluminense \\ Av.~Gal.~Milton Tavares de Souza, S/N, 24210-346, Gragoatá, Niter\'{o}i, Rio de Janeiro, Brazil 
          }

\abstract{
Understanding the applicability of fluid-dynamical models to describe the hot and dense matter produced in the early stages of hadronic collisions is a fundamental problem in the field. In particular, it is not clear to what degree this hydrodynamization process requires proximity to a local equilibrium state. In this contribution, we study this problem in kinetic theory considering an ultrarelativistic gas undergoing strong longitudinal expansion, assuming Bjorken flow. We solve the Boltzmann equation and verify that the system displays considerable deviations from local equilibrium, even though the energy-momentum tensor is well described by fluid dynamics. We further quantify this effect computing the emission of photons in the quark-gluon plasma and verify whether this deviation from equilibrium can be observed.
}
\maketitle
\section{Introduction}
\label{intro}

The main goal of ultrarelativistic heavy-ion collisions is to study the thermodynamic and transport properties of quantum chromodynamics in a controlled environment. After the two nuclei collide, the resulting system evolves into a plasma of asymptotically free quarks and gluons, the quark-gluon plasma (QGP), whose evolution is governed by relativistic dissipative fluid dynamics. However, if and how this system thermalizes -- and, in particular, how this thermalization process occurs so fast -- is still an open question. In particular, it is not trivial to understand how the competition between the system's violent expansion and the interactions between the particles ultimately lead it to a hydrodynamic regime. We approach this problem considering a classical ultrarelativistic gas undergoing a boost-invariant longitudinal expansion, i.e.~Bjorken flow. We determine how well the hydrodynamic approximation for the single-particle distribution function captures the exact solution of the Boltzmann equation. We then estimate how these discrepancies affect thermal photon emissions at early stages of heavy-ion collisions \cite{Shen:2014nfa}, which can be computed in the context of relativistic kinetic theory using the resummation techniques developed in Ref.~\cite{deBrito:2024qow}. 

\section{Kinetic Theory in Bjorken flow}
\label{sec:kin-theo}

In this work, we consider a system of massless classical particles undergoing a longitudinal expansion, assuming Bjorken flow. In this case, the Boltzmann equation assumes a rather simple form -- in particular, employing the relaxation time approximation for the collision term, it becomes
\begin{equation}
\label{eq:boltz-eq-bjorken}
\partial_\tau f_{\mathbf{k}} + \frac{1}{\tau_R} f_{\mathbf{k}} = \frac{1}{\tau_R} f_{0 \mathbf{k}},
\end{equation}
where $\tau_R$ is the relaxation time, $f_{0 \mathbf{k}} = \exp (\alpha - k_0/T)$ is the Boltzmann equilibrium distribution function, with $\alpha$ being the thermal potential, $T$ being the temperature and $k_0$ the particle's energy. For the sake of simplicity, in what follows, we assume a constant relaxation time, $\tau_R = 1$ fm.

Following the method of moments \cite{Denicol:2012cn}, the Boltzmann equation is solved by reconstructing the single-particle distribution from its moments,
\begin{equation}
\label{eq:moment-expansion}
f_{\mathbf{k}}
= 
f_{0\mathbf{k}} \sum_{\ell = 0}^\infty (4\ell + 1) \left( \frac{k_0}{T} \right)^{2\ell}  P_{2\ell}(\cos\Theta) \sum_{n = 0}^\infty  n!  L_n^{(4\ell + 1)} \left( \frac{k_0}{T} \right) \sum_{m=0}^n \frac{(-1)^m \, (m + 2\ell + 1)!}{(n-m)! (m + 4\ell + 1)! m!} \frac{\varrho_{m+2\ell, \ell}}{\varrho_{m+2\ell, 0}^{\mathrm{eq}}},
\end{equation}
where $\cos\Theta = k_{\eta_s}/(\tau k_0)$ and with $P_\ell$ and $L_n^{(m)}$ denoting the Legendre and associated Laguerre polynomials, respectively. 
Furthermore, we have defined the irreducible moments of the single-particle distribution function, $\varrho_{n, \ell}$,
\begin{equation}
    \label{eq:def-moments}
    \varrho_{n, \ell} = \int dK \, k_0^n \, P_{2 \ell} (\cos \Theta) \, f_{\mathbf{k}},
\end{equation}
with $dK = d^3 \mathbf{k}/[(2 \pi)^3 k^0]$ being the Lorentz invariant volume element in momentum space. 

The moments $\varrho_{n, \ell}$ satisfy the following set of coupled differential equations\footnote{The relaxation time approximation requires the imposition of Landau matching conditions, in which the values of the temperature and chemical potential out of equilibrium are defined so that the particle and energy densities are fixed to their equilibrium values, thus $\varrho_{1,0} \equiv \varrho_{1,0}^{\mathrm{eq}}$ and $\varrho_{2,0} \equiv \varrho_{2,0}^{\mathrm{eq}}$.},
\begin{subequations}
\label{eq:hierarchy-eoms-bjorken}
\begin{align}
\frac{d \varrho_{n, \ell}}{d \tau} + \frac{1}{\tau_R} \left[ \varrho_{n, \ell} - e^\alpha \frac{(n+1)!}{2 \pi^2}T^{n+2} \delta_{\ell 0} \right] + \mathcal{P}_{n, \ell} \frac{\varrho_{n, \ell - 1}}{\tau} + \mathcal{Q}_{n, \ell} \frac{\varrho_{n, \ell}}{\tau} - \mathcal{R}_{n, \ell} \frac{\varrho_{n, \ell + 1}}{\tau} =0, \label{eq:eom-rhos-bjorken} \\
\frac{d \alpha}{d \tau} - \frac{2}{\tau} \frac{\varrho_{2,1}}{\varrho_{2,0}^{\mathrm{eq}}} = 0, 
\, \, \, \, \, \, \, \, \,
\frac{dT}{d \tau} + \frac{T}{3 \tau} \left( 1 + 2 \frac{\varrho_{2,1}}{\varrho_{2,0}^{\mathrm{eq}}} \right) = 0, \label{eq:eom-temperature-bjorken} 
\end{align}
\end{subequations}
where we have introduced the following coefficients,
\begin{align}
\mathcal{P}_{n, \ell} &= 2 \ell \frac{(n + 2\ell ) (2\ell - 1)}{(4\ell + 1) (4\ell -1 )}, \, \, \, \,
\mathcal{Q}_{n, \ell} =  \frac{2 \ell (2\ell + 1) + n (24\ell^2 + 12 \ell - 3)}{3 (4\ell - 1) (4\ell + 3)} + \frac{2}{3}, \\
\mathcal{R}_{n, \ell} &= (n - 2 \ell -1) \frac{(2\ell + 1) (2\ell + 2)}{(4\ell + 1) (4\ell + 3)}. \notag
\end{align}

In the context of the method of moments, the Boltzmann equation is solved by reconstructing the single-particle distribution function from its moments, by solving Eqs.~\eqref{eq:hierarchy-eoms-bjorken} and substituting the results into Eq.~\eqref{eq:moment-expansion}. In particular, a fluid-dynamical limit for the single-particle distribution function can be obtained through a coarse graining in which 
it is completely described in terms of the 14 independent degrees of freedom that appear in the conserved currents. Within this truncation -- commonly referred to as 14-moment approximation -- the moment expansion \eqref{eq:moment-expansion} reduces to 
\begin{equation}
    \label{eq:f-14mom}
    f^{\text{14 moment}}_{\mathbf{k}} 
    = 
    f_{0 \mathbf{k}} \left[ 1 + \frac{1}{4 T^2} k_0^2 P_2 (\cos \Theta) \frac{\varrho_{2,1}}{\varrho_{2,0}^{\mathrm{eq}}} \right].
\end{equation}

An exact solution, on the other hand, is obtained when including an infinite number of moments in the expansion for the single-particle distribution function -- that is, taking the sums in Eq.~\eqref{eq:moment-expansion} to infinity. In practice, this is achieved by including a sufficiently large number of moments such that the solutions do not change appreciably, i.e., truncating the sums over $n$ and $\ell$ at finite values $N$ and $L$. However, it was shown that the moment expansion is, in fact, divergent and physically meaningful results can only be achieved by the means of resummation schemes \cite{deBrito:2024qow}. 

\begin{figure}
    \centering
    \includegraphics[width=0.45\linewidth]{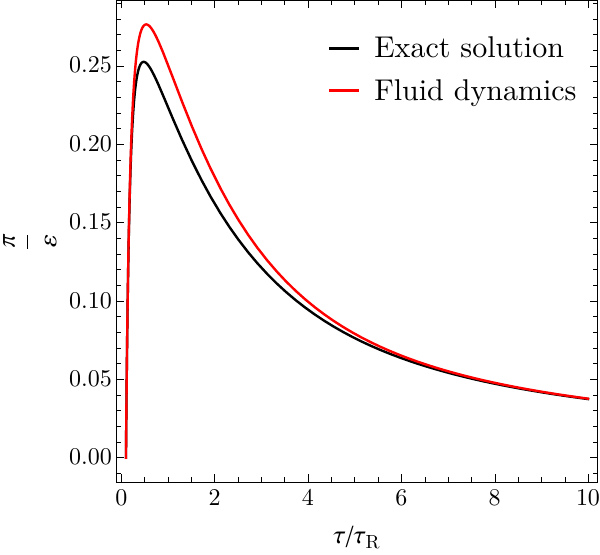} \hfill
    \includegraphics[width=0.45\linewidth]{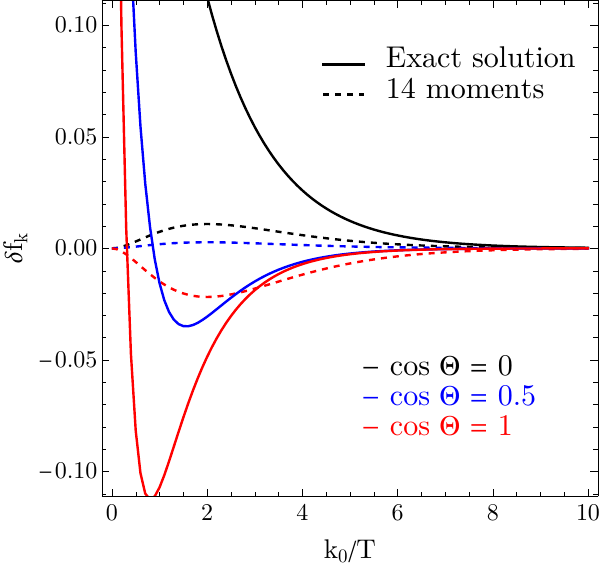}
    \caption{Left panel: Shear-stress rescaled by the energy density as a function of time. Right panel: nonequilibrium single-particle momentum distribution as function of $k_0/T$.}
    \label{fig:hydro-exact-comparison}
\end{figure}

Here we solve the moment equations for $N=20$ and $L=10$, considering an system that is in thermodynamic equilibrium at an initial time $\tau_0 = 0.1$ fm, with a temperature $T (\tau_0) = 10$ GeV, and a thermal potential $\alpha (\tau_0) = 0$. In Fig.~\ref{fig:hydro-exact-comparison}, we compare the exact and hydrodynamic (14-moment) solutions of the Boltzmann equation at time $\tau = 0.5$ fm. 
In the left panel, we display the shear-stress over energy density, $\pi/\varepsilon$, and observe that, even though the system is considerably out of equilibrium, fluid-dynamical solutions provide a good description of the shear-stress tensor. Thus, one might expect that the single-particle distribution function itself is also well described by a fluid-dynamical solution. In order to verify this, in the right panel, we portray the exact (solid curves) and hydrodynamic (dashed curves) solutions for $\delta f_{\mathbf{k}}$ as function of $k_0/T$ considering different values of $\cos\Theta$ and observe that these two solutions are significantly different. As a matter of fact, the 14-moment approximation considerably underestimates the exact solution. 

We thus conclude that the energy-momentum tensor is well captured by hydrodynamics even at the early stages of the collision, whereas the single-particle distribution function is not. In the following, we investigate the impact of this effect in the context of heavy-ion collisions by estimating the emission rate of thermal photons.

\section{Results and discussion}
\label{sec:photon-prod}

We consider the two main two-to-two processes involved in the emission of thermal photons in the quark-gluon plasma, namely quark--anti-quark annihilation ($q \Bar{q} \to \gamma g$) and Compton scattering of quarks and anti-quarks ($q g \to \gamma q$, $\Bar{q} g \to \gamma \Bar{q}$) \cite{Wong:1994}. 
In the forward scattering approximation, the emission rate of photons with momentum $\mathbf{k}$ and energy $k_0$ is given by \cite{Dusling:2009bc}
\begin{equation}
k_0 \frac{d N}{d^3 \mathbf{k}}
=
\frac{5}{9}
\frac{\alpha_{e} \alpha_{s}}{2\pi^{2}}
f_{\textbf{k}}^{(q)}
T^{2} \ln \left( \frac{3.7388 \; k_0}{ g_{\mathrm{s}}^{2} T} \right),
\end{equation}
where $f_{\textbf{k}}^{(q)}$ is the quark distribution function, $\alpha_e$ is the electromagnetic fine-structure constant and $\alpha_s$ is related to the strong coupling, $g_s$, through $\alpha_s = g_s^2/(4 \pi)$. Here, we take $g_s$ such that $\alpha_s = 0.3$. Furthermore, we approximate $f_{\textbf{k}}^{(q)}$ as the distribution calculated in the previous section.

The far from equilibrium momentum distribution function can be quantified by the photon momentum anisotropy, defined as,
\begin{equation}
    \epsilon_{2, \mathbf{k}}
    =
    \frac{\int d\cos\Theta \, \cos (2\Theta) \,
    k_0 \frac{d N}{d^3 \mathbf{k}}}{\int d\cos\Theta \,
    k_0 \frac{d N}{d^3 \mathbf{k}}}.
\end{equation}

\begin{figure}
    \centering
    \includegraphics[width=0.5\linewidth]{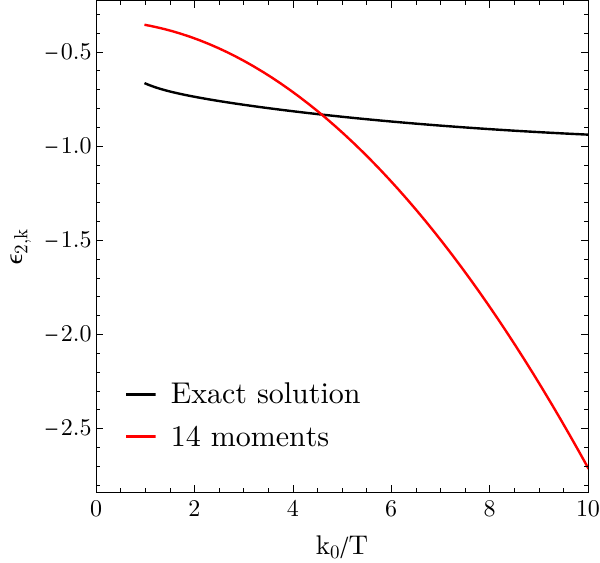}
    \caption{Photon momentum anisotropy as function of $k_0/T$ at time $\tau = 0.8$ fm.}
    \label{fig:photon-aniso}
\end{figure}
Indeed, we see in Fig.~\ref{fig:photon-aniso} that, for small energies, the 14-moment approximation underestimates the exact solution, and later increases with respect to the exact solution for larger values of energy. This happens despite the fact that the energy-momentum is well described by fluid dynamics. Thus, the early time photon emission may be able to tell us if the quark-gluon plasma is indeed close to local equilibrium. 

\section*{Acknowledgments}

The authors thank G.~S.~Rocha and J.-F.~Paquet for helpful discussions. C.~V.~P.~B.~thanks Conselho Nacional de Desenvolvimento Científico e Tecnológico (CNPq) for support, Grant No. 140453/2021-0. G.~S.~D.~also acknowledges CNPq as well as Fundação Carlos Chagas Filho de Amparo à Pesquisa do Estado do Rio de Janeiro (FAPERJ), Grant No.~E-26/202.747/2018.

\end{document}